\documentclass[11pt]{article}
\usepackage{graphicx}
\usepackage{epsfig}
\usepackage{rotate}

\newcommand{\BABARPubYear}    {00}

\newcommand{\BABARConfNumber} {06}
\newcommand{\SLACPubNumber} {8528}

\usepackage{relsize}
\def\babar{\mbox{\slshape B\kern-0.1em{\smaller A}\kern-0.1em
    B\kern-0.1em{\smaller A\kern-0.2em R}}}

\def\epem       {\ensuremath{e^+e^-}}

\def\piz   {\ensuremath{\pi^0}}

\def\pip   {\ensuremath{\pi^+}}
\def\pim   {\ensuremath{\pi^-}}

\def\Kbar  {\kern 0.2em\overline{\kern -0.2em K}{}}

\def\Kp    {\ensuremath{K^+}}
\def\Km    {\ensuremath{K^-}}
 
\def\KL    {\ensuremath{K^0_{\scriptscriptstyle L}}}

\def\Kzb   {\ensuremath{\Kbar^0}}
\def\KzKzb {\ensuremath{K^0 \kern -0.16em \Kzb}}
\def\Dz    {\ensuremath{D^0}}
\def\Dbar  {\kern 0.2em\overline{\kern -0.2em D}{}}

\def\Dzb   {\ensuremath{\Dbar^0}}
\def\DzDzb {\ensuremath{D^0 {\kern -0.16em \Dzb}}}

\def\Bz    {\ensuremath{B^0}}

\def\Bbar  {\kern 0.18em\overline{\kern -0.18em B}{}}

\def\Bzb   {\ensuremath{\Bbar^0}}

\def\BB    {\ensuremath{B\Bbar}} 
\def\BzBzb {\ensuremath{B^0 {\kern -0.16em \Bzb}}}

\mathchardef\Upsilon="7107
\def\Y#1S{\ensuremath{\Upsilon{(#1S)}}}

\def\FourS {\Y4S}

\mathchardef\Deltares="7101
\mathchardef\Xi="7104
\mathchardef\Lambda="7103
\mathchardef\Sigma="7106
\mathchardef\Omega="710A
\def\Deltabar   {\kern 0.25em\overline{\kern -0.25em \Deltares}{}}
\def\Lbar {\kern 0.2em\overline{\kern -0.2em\Lambda\kern 0.05em}\kern-0.05em{}}
\def\Sigbar{\kern 0.2em\overline{\kern -0.2em \Sigma}{}}
\def\Xibar{\kern 0.2em\overline{\kern -0.2em \Xi}{}}
\def\Obar{\kern 0.2em\overline{\kern -0.2em \Omega}{}}
\def\Nbar{\kern 0.2em\overline{\kern -0.2em N}{}}
\def\Xbar{\kern 0.2em\overline{\kern -0.2em X}{}}

\def\pt         {\mbox{$p_T$}}
\def\mes        {\mbox{$m_{\rm ES}$}}

\def\ev   {\ensuremath{\rm \,e\kern -0.08em V}}
\def\kev  {\ensuremath{\rm \,ke\kern -0.08em V}} 
\def\mev  {\ensuremath{\rm \,Me\kern -0.08em V}} 
\def\gev  {\ensuremath{\rm \,Ge\kern -0.08em V}} 
\def\gevc {\ensuremath{{\rm \,Ge\kern -0.08em V\!/}c}} 
\def\tev  {\ensuremath{\rm \,Te\kern -0.08em V}}
\def\mevc {\ensuremath{{\rm \,Me\kern -0.08em V\!/}c}} 
\def\gevcc{\ensuremath{{\rm \,Ge\kern -0.08em V\!/}c^2}} 
\def\mevcc{\ensuremath{{\rm \,Me\kern -0.08em V\!/}c^2}}

\def\cm   {\ensuremath{\rm \,cm}}

\def\mum  {\ensuremath{\,\mu\rm m}} 

\def\invfb   {\ensuremath{\mbox{\,fb}^{-1}}}
\def\mus  {\ensuremath{\rm \,\mus}}

\def\mus        {\ensuremath{\,\mu{\rm s}}}    
\def\gsim{{~\raise.15em\hbox{$>$}\kern-.85em
          \lower.35em\hbox{$\sim$}~}}
\def\lsim{{~\raise.15em\hbox{$<$}\kern-.85em
          \lower.35em\hbox{$\sim$}~}}

\def\ra                 {\ensuremath{\rightarrow}}
\def\to                 {\ensuremath{\rightarrow}}

\def\pep2{PEP-II}

\newcommand{\eqref}[1]{Eq.~(\ref{eq:#1})}

\newcommand{\epjc}      [1]  {{Eur.\ Phys.\ Jour.\ C~{\bf #1}}}

\newcommand{\pl}        [1]  {{Phys.\ Lett.\ {\bf #1}}}      

\newcommand{\prl}       [1]  {{Phys.\ Rev.\ Lett.\ {\bf #1}}} 
\newcommand{\pr}        [1]  {{Phys.\ Rev.\ {\bf #1}}}

\newcommand{\zp}        [1]  {{Z.\ Phys.\ {\bf #1}}}

\def\jetset74   {\mbox{\tt Jetset \hspace{-0.5em}7.\hspace{-0.2em}4}}

\def\Dstarp  {\ensuremath{D^{*+}}}
\def\Dstarm  {\ensuremath{D^{*-}}}

\newcommand{\btodspi}{\ensuremath{\Bz\ra\Dstarm\pi^+}}
\newcommand{\btodsrho}{\ensuremath{\Bz\ra\Dstarm\rho^+}}

\newcommand{\deltae}{\ensuremath{\Delta E}}

\newcommand{\dspi}{\ensuremath{\Dstarm\pip}}
\newcommand{\dsrho}{\ensuremath{\Dstarm\rho^+}}

\long\def\inst#1{\par\nobreak\kern 4pt\nobreak
    {\it #1}\par\vskip 10pt plus 3pt minus 3pt}

\begin{document}
{\pagestyle{empty}

\begin{flushright}
\babar-CONF-\BABARPubYear/\BABARConfNumber \\
SLAC-PUB-\SLACPubNumber
\end{flushright}

\par\vskip 3cm

\begin{center}
\Large \bf Measurement of the branching fractions
for \boldmath $\Bz\to\Dstarm\pip$ and $\Bz\to\Dstarm\rho^+$
\end{center}
\bigskip

\begin{center}
\large The \babar\ Collaboration\\
\mbox{ }\\
July 25, 2000
\end{center}
\bigskip \bigskip

\begin{center}
\large \bf Abstract
\end{center}
Using 5.2 \invfb\ of \epem\ annihilation data recorded with the
\babar\ detector at the \pep2\ storage ring while 
operating on the \FourS\ resonance, a sample of fully reconstructed
\Bz\ decays in the hadronic modes
$\Bz\to\Dstarm\pip$ and $\Bz\to\Dstarm\rho^+$ have been reconstructed.
In this paper, a study of these events is reported, including 
preliminary measurements of the absolute branching
fractions for these modes, which are found to be ${\cal B}(\btodspi)
 = (2.9\pm 0.3\pm 0.3)\times 10^{-3}$ and
${\cal B}(\btodsrho) = (11.2\pm 1.1\pm 2.5)\times 10^{-3}$.

\vfill
\begin{center}
Submitted to the XXX$^{th}$ International 
Conference on High Energy Physics, Osaka, Japan.
\end{center}

\newpage
}

\begin{center}
\small

The \babar\ Collaboration
\bigskip

B.~Aubert,
A.~Boucham,
D.~Boutigny,
I.~De Bonis,
J.~Favier,
J.-M.~Gaillard,
F.~Galeazzi,
A.~Jeremie,
Y.~Karyotakis,
J.~P.~Lees,
P.~Robbe,
V.~Tisserand,
K.~Zachariadou
\inst{Lab de Phys.\ des Particules, F-74941 Annecy-le-Vieux, CEDEX, France}
% ...

\end{center}\newpage

\setcounter{footnote}{0}

\section{Introduction}
\label{sec:Introduction}

The dominant hadronic decay modes of the $B$ meson leading to open charm
in the final state involve tree-level diagrams where the $b\to c$ transition
leads to a charmed meson and an external $W$, which often emerges as a charged $\pi$, $\rho$,
or $a_1$. In this study, we report a new measurement of the branching fractions
for \btodspi\ and \btodsrho~\footnote{Here, 
and throughout this document, we use the convention 
that a particular candidate state also implies the charge conjugate state is included.}. 
The former is known to about 10\% from previous
measurements, 
while the latter has proven more difficult to determine due to the helicity of the 
$\rho$~\cite{ref:CLEO1, ref:CLEO2, ref:ARGUS}. 

\section{The \babar\ detector and dataset}
\label{sec:babar}
The data used in this analysis were collected with the \babar\ detector
while operating
in the \pep2\ storage ring at the Stanford Linear Accelerator Center. 
For this analysis, we
use a sample equivalent to $5.22\pm 0.16$ \invfb\ of integrated luminosity,
collected while running on the \FourS\ resonance. This corresponds to
$(5.93\pm 0.21)\times 10^6$ \BB\ pairs, assuming the \FourS\ decays only
to $B$ mesons. 

The \babar\ detector is described in detail elsewhere~\cite{ref:babar}. Charged
particle tracking is provided by a 5-layer silicon microstrip detector,
capable of stand-alone pattern recognition, and
a 40-layer cylindrical drift chamber.
This is followed outward in radius by 
a specialized particle identification system, based on detection of
Cherenkov light generated in quartz. A calorimeter consisting of
thallium-doped cesium iodide crystals provides detection for photons,
and particle identification for electrons. These devices sit inside the
superconducting coil, which provides a 1.5${\rm \,T}$ field with 3\%
uniformity over the tracking volume. The flux return for the magnet is
finely segmented and instrumented with resistive-plate chambers to
provide both muon identification and crude hadronic calorimetry for
the detection of \KL\ mesons.

\section{\boldmath $B$ reconstruction method}
\label{sec:Analysis}

Hadronic events for this study are selected by a set of simple requirements
designed to minimize systematic errors in $B$ counting. We require more
than three charged tracks forming a primary vertex within 0.5\cm\ of the
beam spot in both transverse directions, 
a sum of charged and neutral energy greater than 5
\gev, and the normalized
second Fox-Wolfram moment for the event, $R_2$, calculated
from charged tracks and neutrals in the \FourS\ frame, less than 0.5. Neutrals
are clusters in the calorimeter with no associated tracks, more than 30\mev\ in energy,
and a lateral moment of the shower distribution less than 0.8.
The $R_2$ requirement is designed to reject
the jet-like continuum events over the more uniformly distributed \FourS\ 
decays.
A fiducial
requirement of $0.410 < \theta_{Lab} < 2.540$ for charged tracks and
$0.410 < \theta_{Lab} < 2.409$ for neutrals,
where $\theta_{Lab}$ is the polar angle to the beam line,
is made in calculating these quantities and
in the subsequent steps of the study.

\Bz\ candidates in the channels \dspi\ and 
\dsrho\ are 
reconstructed using the mode 
$\Dstarm\to\Dzb\pim$, followed by $\Dzb\to\Kp\pim$. The $\rho^+$ is
seen in its decay to $\pip\piz$. All charged tracks are required to
originate close to the
beam spot; the distance of closest approach in the transverse plane,
is required to be less than 1.5 \cm\ and along the beam line
to be less than 3 \cm. The daughters of the \Dzb\ must 
have a transverse momentum,
\pt, greater than 100 \mevc, and include at least 20 drift
chamber hits. The soft pion track is only required to have \pt\ greater
than 70 \mevc, 
taking advantage of stand-alone track finding in the silicon detector to 
improve acceptance. No particle identification is used for this analysis. Neutral
pions are formed from pairs of photons with energy
greater than 30\mev\ and having a lateral shower moment less than 0.8.
The invariant mass of the candidate must lie within $\pm 25$\mevcc\ of the nominal
\piz\ mass.

\Dzb\ candidates are formed from $\Kp\pim$ combinations, where the kaon and pion are
required to have a minimum momentum of 200\mevc, and the invariant
mass of the candidate must be within $\pm 2.5\sigma$ of
the nominal \Dzb\ mass~\cite{ref:pdg98}. All \Dzb\ candidates are required to
have momentum greater than 1.3\gevc\ in the \FourS\ frame. A vertex constraint
fit is performed, for which the $\chi^2$ probability must be greater than 0.1\%.

The reconstructed \Dzb\ is then combined with a soft pion having charge opposite to the kaon
to form \Dstarm\ candidates. A vertex constraint fit is applied to improve the
angular resolution for the soft pion, using a fixed effective vertical beam spot size
of 40\mum. In those cases where the fit converges, \Dstarm\ candidates are selected
by the requirement that $\Delta m = m(\Dzb\pim) - m(\Dzb)$ lies within $\pm 2.5\sigma$
of the nominal mass difference~\cite{ref:pdg98}. The width is taken from a weighted
average of the two-Gaussian distribution that is required to fit the $\Delta m$
distribution. 

\btodspi\ candidates are reconstructed by combining a \Dstarm\ and a \pip\ with 
momentum greater than 500\mevc. The decay \btodspi\ involves a pseudoscalar initial-state particle
decaying into a vector and pseudoscalar, so that the final-state \Dstarm\ is polarized.
Therefore, the helicity angle $\theta_H$ between the soft pion direction and the \Dstarm\
direction in the \Dstarm\ rest frame should be distributed as $\cos^2\theta_H$. In contrast,
combinatorial background is uniform in $\cos\theta_H$. Therefore, $B$ meson candidates are selected with
the additional requirement $|\cos\theta_H|>0.4$.
This requirement removes 40\% of the background and only 5\%
of signal events.

For the \btodsrho\ mode, $\rho^+$ candidates are
formed by combining a \piz\ meson and a charged pion with momentum greater than
200\mevc. We require the momentum of the $\rho^+$ to be greater than 1\gevc\, and
the \pip\piz\ mass to lie within 150\mevcc\ of the $\rho^+$ mass. Finally,
the opening angle, $\theta_T$,
between the thrust angle of the $B$ candidate and the thrust axis
of the remaining charged and neutrals in the event
is required to satisfy $|\cos\theta_T|<0.9$ in order
to remove continuum backgrounds.

In the case of a correctly reconstructed $B$ meson produced by the 
decay of an
\FourS, within the experimental resolution, the measured 
sum of neutral and charged energies, $E^*_m$, must be equal to the 
beam energy, $E^*_b$, both evaluated in the \FourS\ frame.
We define \deltae\ to be the difference between the measured
$B$ candidate energy and beam energy in the \FourS\ frame,
$\Delta E = E^*_m- E^*_b$.
The resolution on $\sigma_{\deltae}$ is predicted to be
$16.0\pm 0.6$\mev\ for $\Dstarm\pi^+$
and 28\mev\ for $\Dstarm\rho^+$.
The beam-energy substituted mass, \mes, is defined as
$  m_{ES}^2  =  
\left(E^*_b\right)^2 -\left(\sum_i \mbox{\boldmath $p$}_i\right)^2$,
where the $\mbox{\boldmath $p$}_i$ is the momentum of the $i$th
daughter of the $B$
candidate.
The predicted resolution in \mes\ is typically about 
2.5 \mevcc\ for \dspi\ and 3.1 \mevcc\ for \dsrho. 
This
is about a factor of 10
better than the resolution in the reconstructed invariant mass. The
resolution for \mes\ is dominated
by the beam energy spread rather than by the detector resolution. It is 
largely uncorrelated with the error on \deltae.
    
The variables \deltae\ and \mes\ are used to define a signal region
and also sideband regions for background studies. For all modes, 
the region between 5.2 and 5.3\gevcc\ in \mes\ and between
$\pm 300$ \mev\ in \deltae\ is used to study the $B$ candidates.
The peak position, $m_B^0$, which should be the nominal $B$ mass, and the
resolution $\sigma_{\mes}$ are extracted
from the distribution of \mes\ after
requiring 
\deltae\ be consistent with zero to within $\pm 2.5 \sigma$.
The resolution in \deltae\ is extracted from the \deltae\ distribution 
obtained by requiring that \mes\ lie within $\pm 2.5 \sigma_{\mes}$ of
$m_B^0$.

The signal region in the two dimensional plane \mes\ versus 
\deltae\ is defined as a area $\pm 2.5 \sigma $ wide 
centered at the nominal $B$ mass, $m_B^0$, and
at $\deltae = 0$. 
By
staying below $|\deltae|= m_{\pi}$, we avoid correlated background
from $B$ decays where a real final-state charged pion is either not 
included in
the reconstruction or a random soft pion from the recoil $B$ is added to 
the observed state.

\begin{figure}[htb]
\begin{center}
\begin{tabular}{lr}
\mbox{\epsfxsize=8.5cm\epsffile{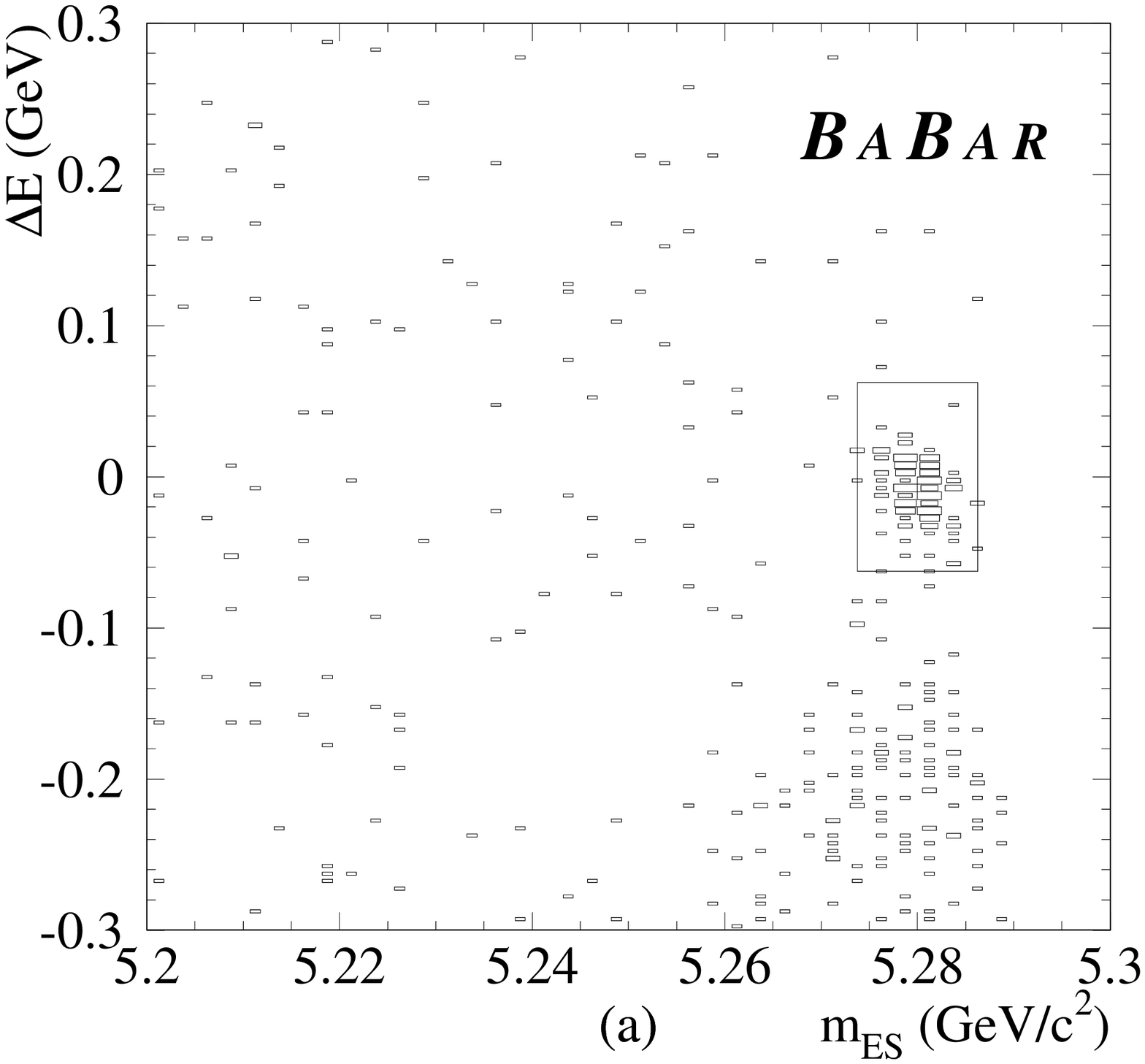}} & 
\mbox{\epsfxsize=8.5cm\epsffile{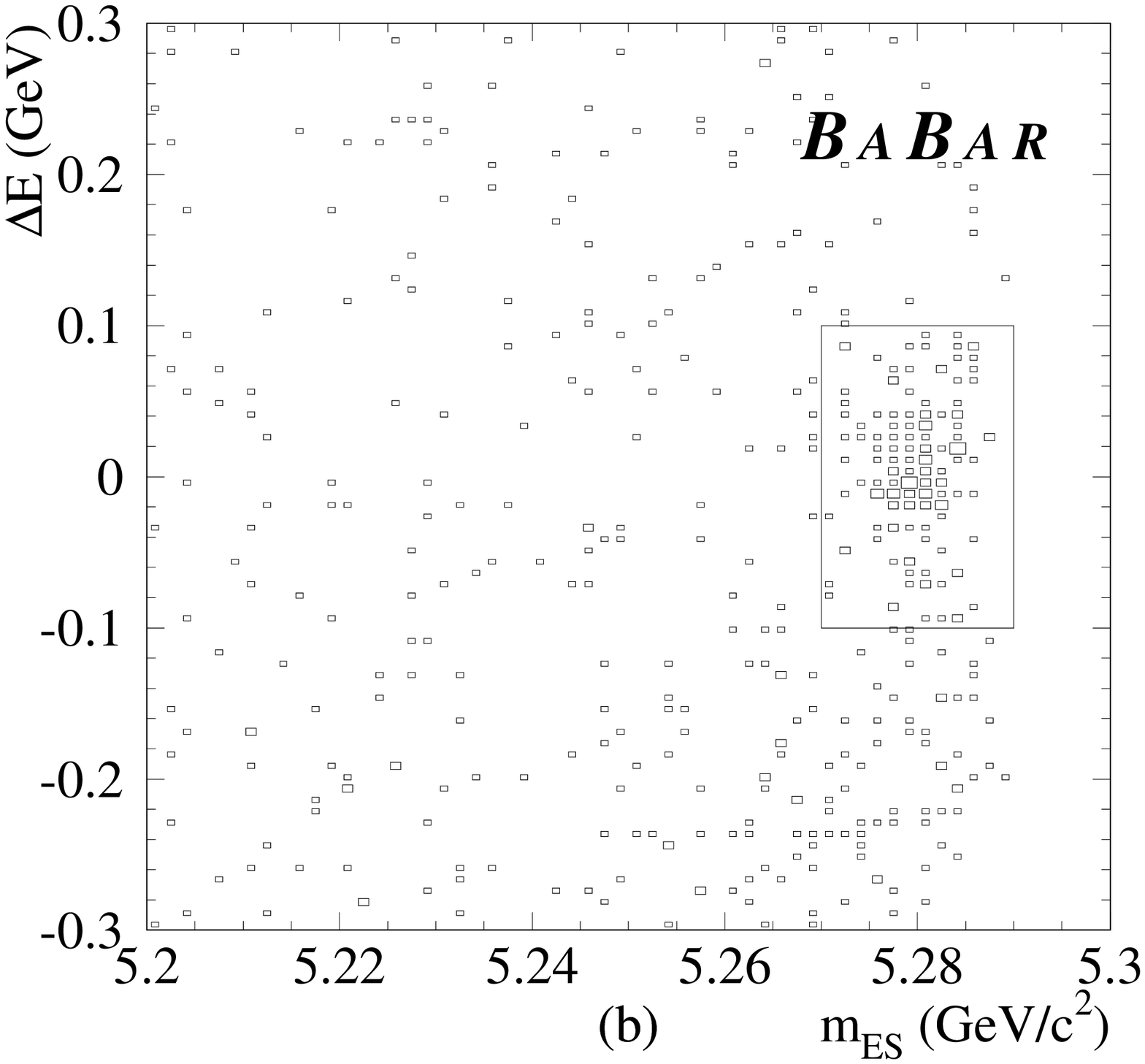}} \\
\end{tabular}
\end{center}
\caption{Distribution in data of \deltae\ versus \mes\ for $B$ candidates in
the channels (a) \btodspi\ and (b) \btodsrho.}
\label{fig:deltae}
\end{figure}

Only one candidate per event is allowed to appear in the \mes\ versus
\deltae\ distribution. The criteria selected is to consider only the entry
with the smallest absolute value for \deltae. The resulting two dimensional
distribution of candidate events in \deltae\ and \mes\ 
is shown in Fig.~\ref{fig:deltae}.

\section{Branching fraction measurements}
\label{sec:Physics}

\begin{figure}
\begin{center}
\begin{tabular}{lr}
\mbox{\epsfxsize=7.5cm\epsffile{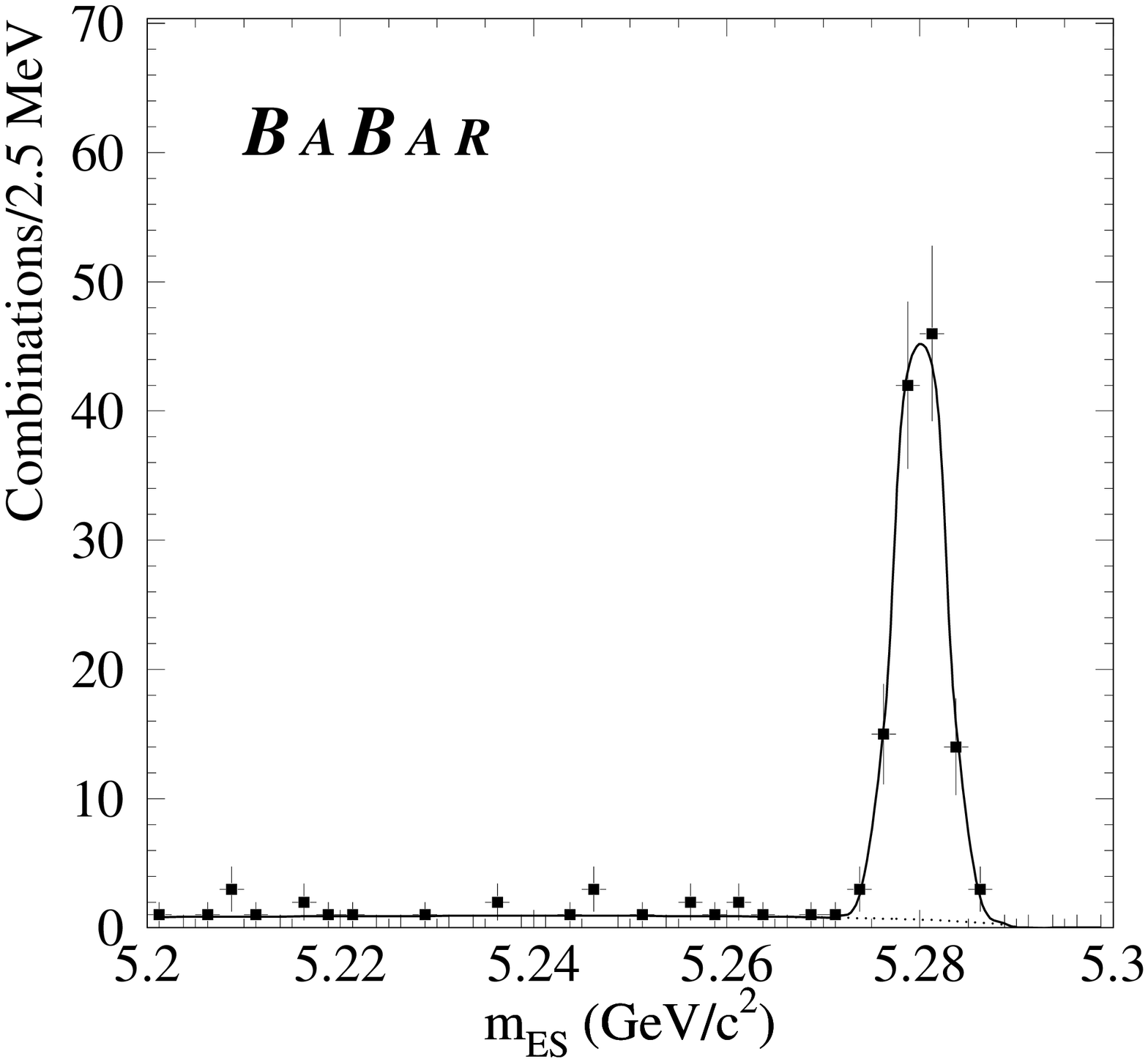}} & 
\mbox{\epsfxsize=7.5cm\epsffile{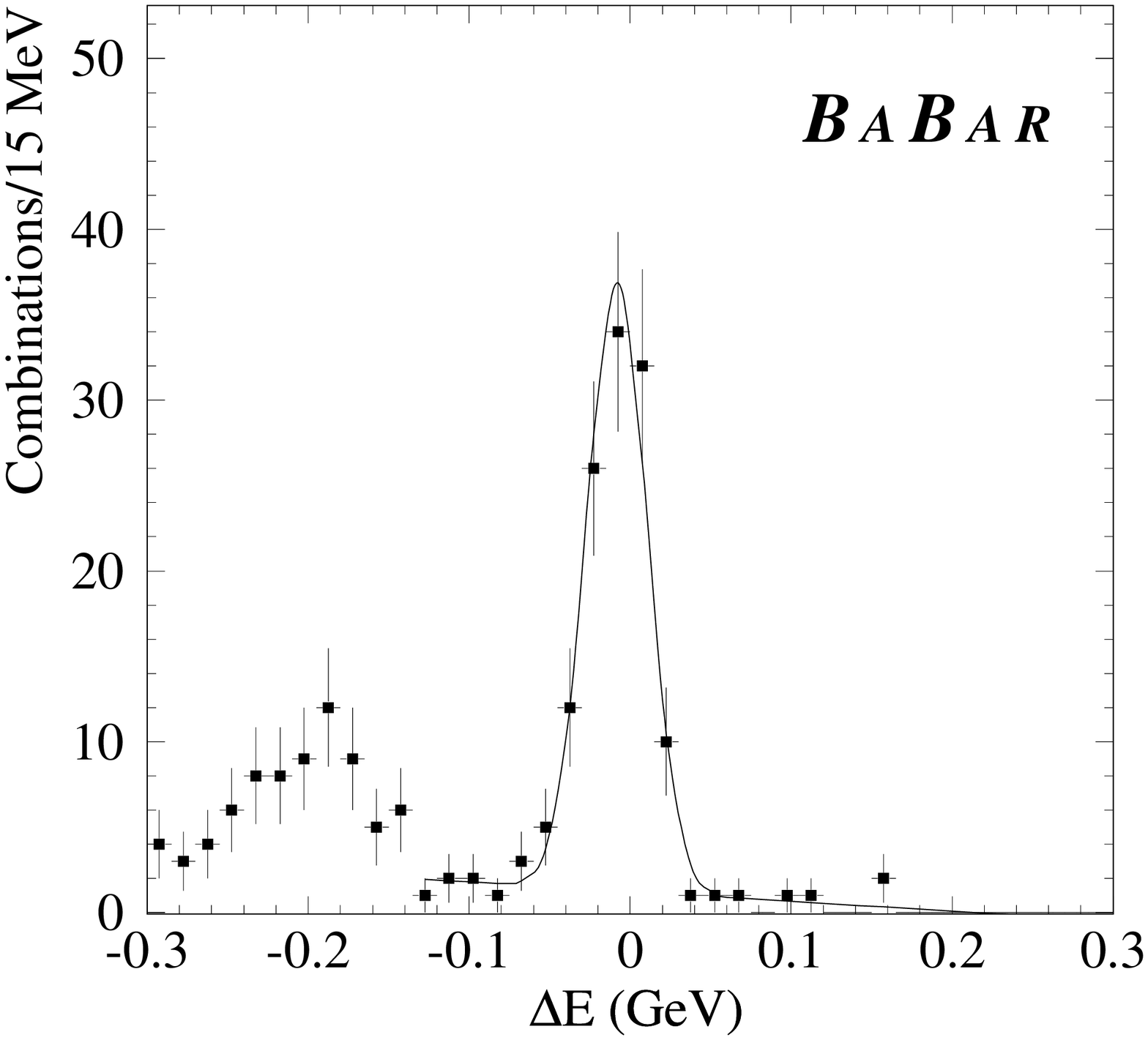}} \\
\end{tabular}
\end{center}
\caption{
Distribution of \mes\ for $| \deltae\ |< 2.5\sigma_{\Delta E}$ (left), and
\deltae\ for $|\mes-m_B^0|< 2.5\sigma_{\mes}$ (right) in the channel \btodspi. The fits are described in the text.}
\label{fig:dspi}
\end{figure}

The measurement of  branching fractions
requires an estimate of the combinatorial background in the
\mes\ distribution near the nominal $B$ mass. 
For the channel \btodspi,
the \mes\ distribution, shown in 
Fig.~\ref{fig:dspi},  
is fitted with a 
background function~\cite{ref:ARGUS} which parameterizes how the phase space approaches zero
as the energy approaches $E^*_b$:  
\begin{eqnarray}
f_{BG}(\xi) &=& N \xi\sqrt{1-\xi^2}\exp(\kappa(1-\xi^2))
\end{eqnarray}
where $\xi=\mes/E^*_b$, and 
the normalization and the shape are determined by the
parameters $N$ and $\kappa$. 
The signal is characterized by a Gaussian distribution with free mass, $m_B$, and width, $\sigma_{\mes}$. 
The projection of the signal as a function of \deltae, also shown in 
Fig.~\ref{fig:dspi}, is
fitted using
a linear background function plus a single Gaussian distribution
with free mean and width, $\sigma_{\deltae}$. The region $-0.3<\deltae<-0.13$\gev, containing feeddown from
$\Dstarm\rho$ where a pion is left out of the reconstruction, is
excluded from the fit. The observed width of the signal in \mes\ is $2.45\pm 0.18$\mevcc\ while the
\deltae\ resolution is observed to be $18.8\pm 0.9$\mev.
Acceptances are calculated in terms of the fitted
widths observed in data and Monte Carlo simulation.

\begin{figure}[htb]
\begin{center}
\begin{tabular}{lr}
\mbox{\epsfxsize=7.5cm\epsffile{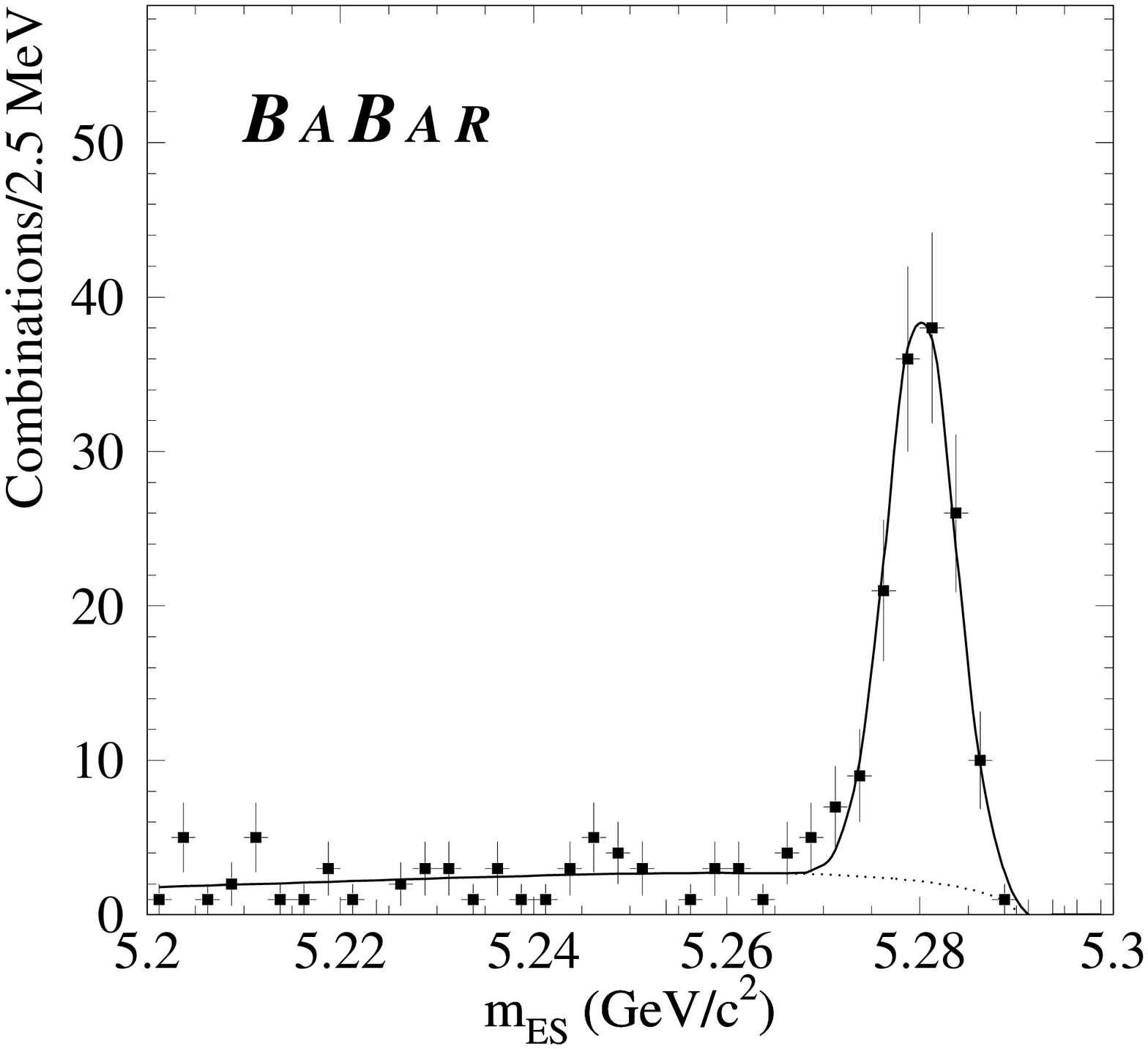}} & 
\mbox{\epsfxsize=7.5cm\epsffile{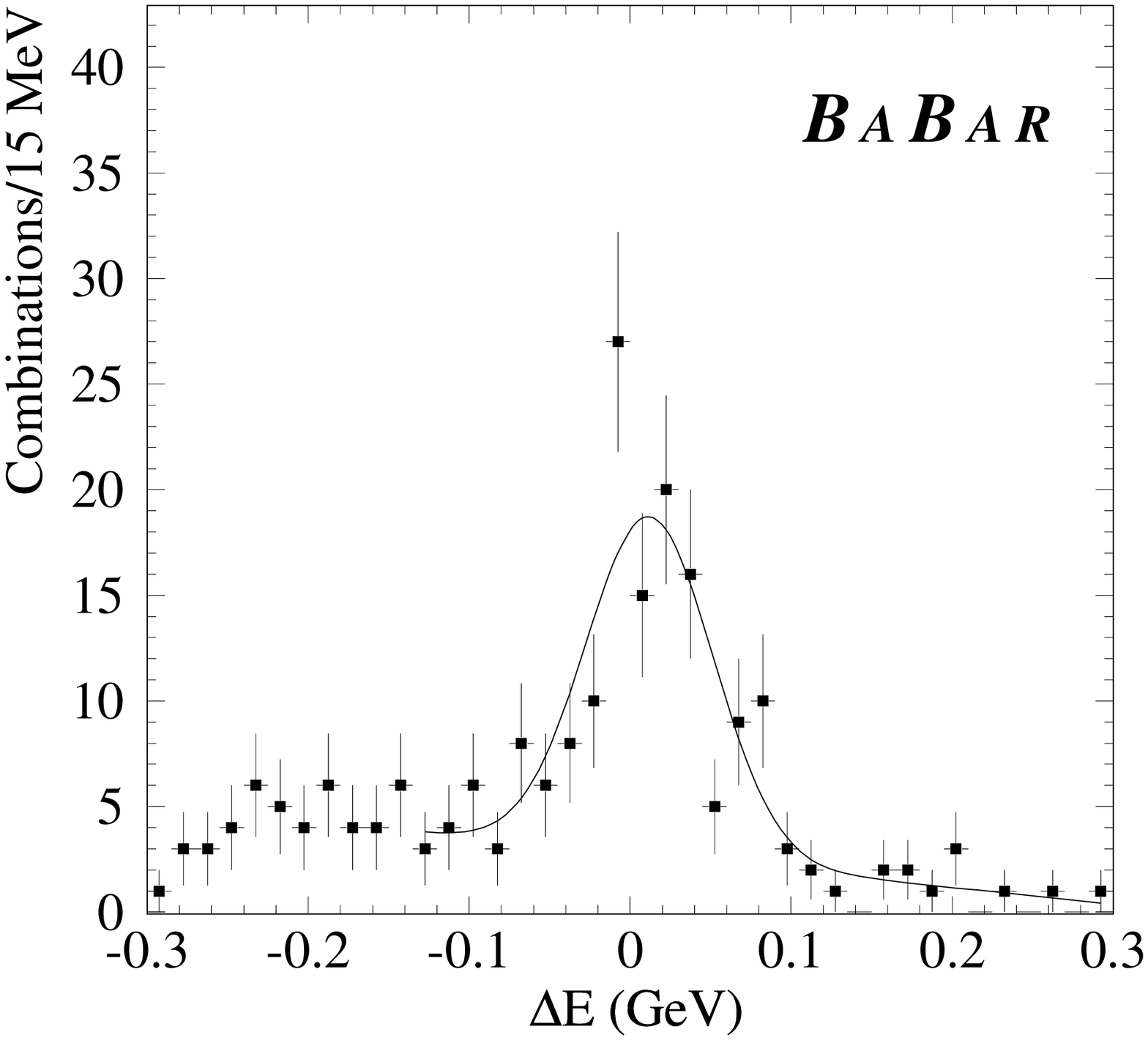}} \\
\end{tabular}
\end{center}
\caption{
Distribution of \mes\ for $| \deltae\ |< 2.5\sigma_{\Delta E}$ (left), and
\deltae\ for $|\mes-m_B^0|< 2.5\sigma_{\mes}$ (right) in the channel \btodsrho. The fits 
are described in the text.}
\label{fig:dsrho}
\end{figure}

For the channel, \btodsrho\,
the observed \mes\ distribution is shown in 
Fig.~\ref{fig:dsrho}. This has been fitted with  
the same background function and signal function
described above.
The projection of the signal as a function of \deltae, also shown in 
Fig.~\ref{fig:dsrho}, is
fitted using
a linear background function plus a single Gaussian distribution
with free mean and width, $\sigma_{\deltae}$,
with the region $-0.3<\deltae<-0.13$\gev\ once more excluded. The width of 
the signal in \mes\ is $3.5\pm 0.3$\mevcc\ while the
\deltae\ resolution is observed to be $39.5\pm 4.7$\mev.

\begin{table}
\caption{Observed and expected yields and efficiencies for \Bz\ decay modes.}
\vspace{0.3cm}
\begin{center}
\label{tab:results}
\begin{tabular}{|l|c|c|} \hline
\Bz\ mode & Observed yield & Efficiency (\%) \\ \hline\hline
$\Dstarm\pip$   & $119\pm 11 $ & $27.0\pm 1.0$   \\
$\Dstarm\rho^+$ & $131\pm 13$  & $7.6\pm  0.6$   \\
\hline
\end{tabular}
\end{center}
\end{table}

A detailed Monte Carlo simulation is used to determine the acceptance for the signal events. Control
samples are used to determine the uncertainties on crucial performance characteristics of the
simulation. The fitted yields for signal events and
estimated efficiencies for each channel are listed in Table~\ref{tab:results}.

A variety of sources contribute to the systematic errors on the final branching fraction results. The
number of produced $B$ mesons is extracted from the ratio of multihadron-to-muon pair events on- and
off-resonance, after application of the simple event selection criteria described above. After
extrapolation from accepted to produced numbers of \BB\ pairs, the estimated uncertainty is 
3.6\%~\cite{ref:babar}.

The primary check on charged track efficiency is obtained by
studying the probability for observing drift chamber versus silicon detector-only tracks in
inclusive $\Dz\to\Km\pip\pip\pim$. This is compared with the rate for finding the third track 
in one-versus-three topology
tau-pair decays. A final check is the observed multiplicity distribution in \FourS\ events.
A systematic error of 2.5\% per track
with $\pt>1$\gevc\ on the overall efficiency scale is determined by comparing the three methods.
The soft pion efficiency is determined from
a study of the forward-backward asymmetry in observed \Dstarm\ decays, and is also constrained
to some extent by the same charm and tau-pair studies. 
The \pt\
resolution in the Monte Carlo events is adjusted to reproduce the resolution seen in cosmic
ray muons. The overall systematic uncertainty in the final result due to tracking efficiency is estimated
to be 7.9\% for the final states considered in this study.
An additional systematic error contribution come from changes in the ratio of observed and predicted 
efficiencies as the
selection requirements are varied within a reasonable range (3\%).

In the case of the \btodsrho\ channel, there are additional systematic errors due to the $\rho^+$ and 
its polarization.
The reconstruction efficiency for the \piz, as modeled in our Monte Carlo simulation, is verified to be
 accurate
to within 5\%~\cite{ref:babar}. The width of the signal in the \deltae\ projection is modestly larger than
 predicted by the Monte Carlo simulation.
Therefore, we have varied the \deltae\ requirement for the signal band over a wide range around the 
nominal 2.5$\sigma$,
and estimate a further 10\% uncertainty due to this requirement.
Finally, we have extracted the branching 
fraction separately
in the two hemispheres of the $\rho$ helicity angle, $\theta_H(\rho)$, greater and less than zero.
 A further 15\% systematic 
error is assigned on the basis of the observed difference in branching fraction. The helicity distributions
for the $\rho$ and the \Dstarm\ are consistent with expectations based on previous measurements~\cite{ref:CLEO2}
and our Monte Carlo efficiency calculation, although we have not yet attempted to determine the polarization
from the data. 

\begin{figure}[htb]
\begin{center}
\mbox{\epsfxsize=8.5cm\epsffile{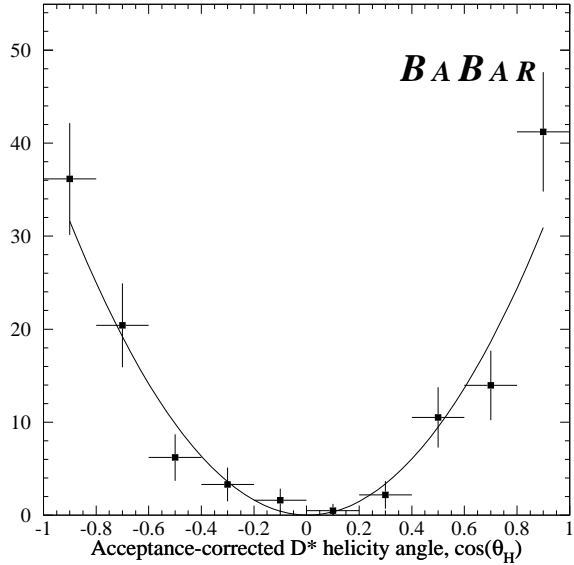}}
\end{center}
\caption{Distribution of \btodspi\ as a function of the helicity angle of the soft pion, $\cos\theta_H$, in
the \Dstarm\ rest frame after acceptance correction. The vertical axis has arbitrary units.}
\label{fig:helicity}
\end{figure}

An additional check of the result for \btodspi\ is to relax the $\cos\theta_H$ requirement, 
and extract the fitted signal as a function
of $\cos\theta_H$ instead. The distribution of the $B$ candidates within
$\pm 2.5\sigma_{\mes}$ of the nominal $B$ mass
and $\pm 2.5\sigma_{\deltae}$ of $\deltae=0$ exhibits no forward-backward
asymmetry after acceptance corrections
as can be seen in Figure~\ref{fig:helicity}. This confirms our understanding of the soft pion efficiency in the 
\Dstarm\ decay. Overlayed on the 126 signal candidates, with an estimated background of 4 events, is a fit with
$\cos^2\theta_H$, yielding a $\chi^2$ for goodness-of-fit of 8.6
for 7 degrees of freedom. 

The PDG compilation~\cite{ref:pdg98} of measured branching fractions for the $\Dstarp\to\Dz\pip$, 
$(68.3\pm 1.4)$\%, and
$\Dz\to\Km\pip$, $(3.83\pm 0.09)$\%, are used in computing our final results. The measurement errors 
on these branching
fractions are included in the systematic error on our final result. We also assume the \FourS\ decays into
\Bz\Bzb\ pairs with a 50\% fraction; no systematic error is assigned to this value.

Based on fitted yield of signal events, the estimated efficiency, and the number of produced $B$ mesons 
in our sample,
the preliminary results for the branching fractions for \btodspi\ and \btodsrho\ 
are $(2.9\pm 0.3\pm 0.3)\times 10^{-3}$ and
$(11.2\pm 1.1\pm 2.5)\times 10^{-3}$ respectively. 
The branching fraction for \btodsrho\ includes all non-resonant and quasi-two-body contributions that 
lead to a \pip\piz\ invariant mass
in the $\rho$ band. However, the acceptance for non-resonant \Dstarm\pip\piz\ decays
is about 15\% of $\Dstarm\rho^+$ so that, combined with the known branching fraction for this mode, 
the non-resonant contribution to our result
for \btodsrho\ is expected to be quite small. 
Both branching fraction results compare well with previous measurements and with the world
average~\cite{ref:pdg98}. 

\section{Summary}
\label{sec:Summary}

\Bz\ decays to \Dstarm\pip\ and $\Dstarm\rho^+$ have been studied using the decay chain 
$\Dstarm\to\Dzb\pim$, followed by $\Dzb\to\Kp\pim$. The preliminary
branching fractions obtained for these channels,
$(2.9\pm 0.3\pm 0.3)\times 10^{-3}$ and
$(11.2\pm 1.1\pm 2.5)\times 10^{-3}$ respectively, are compatible with previous 
observations~\cite{ref:CLEO1, ref:CLEO2, ref:ARGUS}.

\section{Acknowledgments}
\label{sec:Acknowledgments}

We are grateful for the contributions of our \pep2\ colleagues in
achieving the excellent luminosity and machine conditions
that have made this work possible.
We acknowledge support from the
Natural Sciences and Engineering Research Council (Canada),
Institute of High Energy Physics (China),
Commissariat \`a l'Energie Atomique and
Institut National de Physique Nucl\'eaire et de Physique des Particules
(France),
Bundesministerium f\"ur Bildung und Forschung
(Germany),
Istituto Nazionale di Fisica Nucleare (Italy),
The Research Council of Norway,
Ministry of Science and Technology of the Russian Federation,
Particle Physics and Astronomy Research Council (United Kingdom), the
Department of Energy (US),
and the National Science Foundation (US). In addition, individual support 
has been received from the Swiss 
National Foundation, the A. P. Sloan Foundation, the Research Corporation,
and the Alexander von Humboldt Foundation.
The visiting groups wish to thank 
SLAC for the support and kind hospitality
extended to them.

\end{document}